\documentclass[12pt]{article}
\usepackage{amssymb}
\usepackage{amsfonts}

\def\beq{\begin{equation}}
\def\eeq{\end{equation}}
\def\bb{\begin{eqnarray}}
\def\ee{\end{eqnarray}}

\begin{document}
\begin{titlepage}

\begin{center}
{\Large\bf Ghost Constraints on Modified Gravity}
\end{center}
\vspace{0.5cm}
\begin{center}

{\large Alvaro N\'u\~nez\footnote{e-mail address:
an313@scires.nyu.edu}  and Slava Solganik\footnote{e-mail address:
ss706@scires.nyu.edu}}
\vspace{0.5cm}

{\em New York University, Department of Physics, New York, NY 10003, USA}.\\
\end{center}

\hspace{1.0cm}

\begin{abstract} 
We show that general infrared modifications of the Einstein-Hilbert action obtained by addition of curvature invariants are not viable.
These modifications contain either ghosts or light gravity scalars. A very specific fine-tuning might solve the problem of ghosts, but the resulting
theory is still equivalent to a scalar-tensor gravity and thus gives a corrupted picture of gravity at the solar system scale. The only known loophole
is that the theory becomes higher dimensional at large distances. The infinite number of degrees of freedom introduced in this way is not reducible to
the addition of an arbitrary function of curvature invariants. 
\end{abstract}

\end{titlepage} 

\newpage 

\bigskip 

The observed current accelerated expansion of the universe \cite{Riess:1998cb}-\cite{Tonry:2003zg} is an interesting and perplexing problem that has
become a serious challenge for gravity theory. As a result, a considerable effort to modify gravity has been made in order to accommodate this large
distance observation. The modifications, however, should not contradict solar system gravity experiments, as was pointed out in \cite{Dvali:2002vf},
\cite{Dvali:2004ph}.

The attempts to solve the issue have been based in a variety of approaches. Among all of them we distinguish between those with extra dimensions and
those in which the modification occurs strictly in terms of the 4D action by means of introducing local curvature invariants. The key difference
between these two approaches is that extra dimensions add an infinite number of degrees of freedom to the effective 4D theory.  These degrees of
freedom can sum up into a pathology free theory.

An illustrative example is the Kaluza-Klein theory. A 5D gravity with a compact extra dimension is equivalent to an effective 4D theory with an
infinite number of massive spin-2 degrees of freedom, the sum over the tower of Kaluza-Klein modes.  As soon as one truncates the series,
inconsistencies in the theory are produced. Truncation explicitly breaks higher dimensional covariance. The lesson one learns is that it is barely
possible to consistently modify a 4D action by adding a finite number of degrees of freedom, even with an infinite number of higher derivative terms.
The cancellation of inconsistencies occurs only when one adds an infinite number of degrees of freedom, like in the case of theories with extra
dimensions. A non contradictory scheme of this kind was suggested in the work \cite{Deffayet:2001pu}, the authors of which developed it further in the
context of the proposed in \cite{Dvali:2000hr} model.

The purpose of the present paper is to formally show that under very general assumptions the approaches based on addition of invariants fail. 
We do this by calculating the propagator for a generic theory of this kind. 

The attractively simple modification of gravity which takes into account a generic dependence of the action on the curvature scalar
$\sqrt{-g}R\rightarrow\sqrt{-g}f(R)$ was widely discussed in the infrared regime \cite{Carroll:2003wy}. However, this modification reduces essentially
to the theory of Einstein's gravity plus that of an extra scalar field. On the language of conformal rescaling this was shown
in~\cite{Barrow:xh},~\cite{Kalara:1990ar}. Recently we have proved it by explicitly calculating the propagator~\cite{Nunez:2004ji}. The scalar has to
be light to influence the infrared behavior of the theory. This cannot accommodate both the accelerated expansion and the solar system experiments.

One may wonder, what will happen if we make a more general modification? We can make the action dependent on other scalar invariants,
$\sqrt{-g}R\rightarrow\sqrt{-g}f(R,R_{\mu\nu}R^{\mu\nu},R_{\mu\nu\rho\lambda}R^{\mu\nu\rho\lambda})$. Some cosmological applications of these models
were considered in~\cite{Carroll:2004de}, the authors of which did not discuss the possible appearance of ghosts in the hope that this problem either
will not raise in these models, or it will be solved by some unknown mechanism. We show that changing the action by replacing the scalar curvature by a
function of the invariants $R$, $R_{\mu\nu}R^{\mu\nu}$, $R_{\mu\nu\rho\lambda}R^{\mu\nu\rho\lambda}$ does lead to the appearance of ghost
instabilities.

Summarizing, we can say that as long as such a theory is weakly coupled, it is inconsistent due to negative norm states or because of its contradiction
with solar system experiments. The reason is that the Einstein-Hilbert action altered by terms depending on curvature invariants introduces a finite
number of poles which correspond either to ghosts or to light scalars. The situation in extra-dimensional theories is substantially different.  In
particular, the DGP model~\cite{Dvali:2000hr} satisfactory modifies gravity in the infrared~\cite{Dvali:2003rk} and is ghost free.  Instead of the
poles in the momentum space the propagator has a branch cut. This branch cut cannot be formed by the poles appearing when one adds a countable number
of terms to the original 4D action.

We start from a generic theory with the action 
\bb 
\label{action} S=\int d^4x\,\sqrt{-g}
f\left(R,P,Q\right)+ S_{matter} 
\ee 
where $P\equiv R_{\mu\nu}R^{\mu\nu}$, $Q\equiv R_{\mu\nu\rho\lambda}R^{\mu\nu\rho\lambda}$, $f(R,P,Q)$ is some function  and $S_{matter}$ 
is the action for the matter fields. We will show that these more generic actions suffer from even worse problems than the $f(R)$ one. In particular
they contain a spin-2 ghost, which makes them unphysical.

To analyze the theory (\ref{action}) we will consider it from the point of view of propagating degrees of freedom by deriving the particle propagator.  
In this way one can easily see whether there are ghosts, tachyons and what the spins of these states are.

The variation of  the action with respect to the metric leads to the equations of motion
\bb
\label{eq}
&\:& \left(\nabla_{\mu}\nabla_{\nu}-g_{\mu\nu}\nabla^{\alpha}\nabla_{\alpha}-R_{\mu\nu}\right)f_R +\nonumber\\ &+&
\left(2\nabla_{\mu}\nabla^{\alpha}R_{\alpha\nu}-\nabla_{\alpha}\nabla^{\alpha}R_{\mu\nu} -g_{\mu\nu}\nabla^{\alpha}\nabla^{\beta}
R_{\alpha\beta}-2 R_{\alpha\mu}R^{\alpha}_{\nu}\right)f_P+\nonumber\\
&+&\left(4\nabla^{\beta}\nabla^{\alpha}R_{\alpha\mu\nu\beta}
-2 R_{\alpha\beta\gamma\mu}R^{\alpha\beta\gamma}_{\;\;\;\;\;\;\;\nu}\right)f_Q +\frac{1}{2}f g_{\mu\nu}=T_{\mu\nu},
\ee
where $\nabla_{\alpha}$ is a covariant derivative and we use the notation
\bb
f\equiv f(R,P,Q),\;\;\; f_R\equiv \frac{\partial f}{\partial R},
\;\;\; f_P\equiv \frac{\partial f}{\partial P}
,\;\;\; f_Q\equiv \frac{\partial f}{\partial Q},\;\;\;
f_{RR}\equiv \frac{\partial^2 f}{\partial R^2},\nonumber\\
\;\;\; f_{PP}\equiv \frac{\partial^2 f}{\partial P^2}
,\;\;\; f_{QQ}\equiv \frac{\partial^2 f}{\partial Q^2}
,\;\;\; f_{RQ}\equiv \frac{\partial^2 f}{\partial R\partial Q}
,\;\;\; f_{RP}\equiv \frac{\partial^2 f}{\partial R\partial P}.
\ee
These equations have a constant curvature $R=R_0 ={\rm const}$ maximally symmetric solution in vacuum,
implying
\bb
R_{\lambda\mu\nu\sigma}=\frac{R_0}{12}(g_{\lambda\nu}g_{\mu\sigma}-g_{\lambda\sigma}g_{\mu\nu})
\ee
and
\bb
R_{\mu\nu}=\frac{1}{4}R_0g_{\mu\nu}.
\ee
The curvature $R_{0}$ is defined by the equation
\bb
f -\frac{1}{2}R_0 f_R-\frac{1}{4}R_0^2 f_P-\frac{1}{6}R_0^2 f_Q=0,
\label{curvature}
\ee
where $f, f_R, f_P, f_Q$ should be taken at the point $R=R_0$. The other two invariants are then $Q=R^2_0/4$ and $P=R^2_0/6$.

We want to linearize the equations of motion on the maximally symmetric background solution to get the propagator.  We take the metric in the form
$g_{\mu\nu}=g_{\mu\nu}^{(0)}+h_{\mu\nu}$, where $g_{\mu\nu}^{(0)}$ is the solution of (\ref{eq})  corresponding to our constant scalar curvature $R_0$,
next expand the equations of motion up to linear order terms and find the propagator. As a result we obtain the equation
\bb
O_{\alpha\beta\mu\nu}h^{\mu\nu}=T_{\alpha\beta},
\ee
where $O_{\alpha\beta\mu\nu}$ is the inverse propagator. The explicit form of this operator is given in Appendix A.

In order to find the inverse of the above operator and get the propagator, it is convenient to use spin projectors
$P^2$, $P^1_m$, $P^1_e$, $P^1_b$, $P^1_{me}$, $P^1_{em}$, $P^0_s$, $P^0_w$, $P^0_{sw}$, $P^0_{ws}$ 
\cite{VanNieuwenhuizen:fi}. The explicit form of the projectors is given in Appendix B.

The key point will be to notice that the higher than second derivative terms give rise to ghosts. For example, if the inverse propagator has the form
\bb
(\partial^4 +...)P^2+...\equiv(\partial^2 +A)(\partial^2 +B )P^2 +...,
\ee  
the propagator is
\bb
G=\frac{1}{A-B} \left( \frac{1}{\partial^2+B} - \frac{1}{\partial^2+A} \right)P^2 +....
\ee
This propagator has obviously a ghost. We want to stress that the situation here is distinct from that discussed in \cite{Nunez:2004ji} for $f(R)$ gravity. There
the ghost related to the $P^0_s$ projector was actually necessary for the correct cancellation of the longitudinal part of the $P^2$ projector. Here we get a 
real ghost. It also can be shown that even when $A=B$ there is a ghost, and it is impossible to make a consistent theory.

In the light of the above discussion we can analyze the propagator arising from $O_{\alpha\beta\mu\nu}$. Since the operator is written in terms of 
covariant derivatives it is not straightforward to write the inverse. However, what we must be concerned with are the highest derivatives, one can 
notice that
\bb
\nabla_{\alpha}\nabla_{\beta}\nabla_{\mu}\nabla_{\nu}=
\partial_{\alpha}\partial_{\beta}\partial_{\mu}\partial_{\nu}+(C\partial\partial\partial)_{\alpha\beta\mu\nu}+...,
\ee
Where $C$ is a curvature depending coefficient. Thus we can write the propagator in the form
\bb
G=\frac{P^2}{\left(f_P/2 +2f_Q\right)\partial^4+ A \partial^2+ ...}+...
\ee
This means that under general assumptions the theory has ghost. The only possibility not to have a graviton ghost is for the curvature $R_0$ to satisfy 
the condition 
\bb
\left(f_P/2 +2f_Q\right)|_{R=R_0}=0.
\label{condition}
\ee
The realization of this requires the two algebraic equations (\ref{curvature}) and (\ref{condition}) to have the same solution $R=R_0$, which is
not trivial to fulfill. Moreover, even in this case it is not guaranteed that the scalar field ($P^0_s$ projector) will not be a ghost. In the best case
scenario we will get the theory reduced to a scalar-tensor gravity \cite{Nunez:2004ji}.

We further give the explicit form of the propagator for a flat background ($R=0$),
\bb
G=- \frac{P^2}{\left(f_P/2 +2f_Q\right)\partial^4+f_R/2 \partial^2+f/2}-\nonumber\\
- \frac{P^0_s}{2\left(f_P+f_Q+f_{RR}\right)\partial^4-f_R \partial^2-f/2}
-\frac{P^1_m}{f/2}-\frac{P^0_w}{f/2}
\ee
This propagator has a ghost in the spin-2 projector $P^2$ term. Moreover, we also expect a ghost in the scalar degree of freedom $P^0_s$, which got 
modified compared to the $f(R)$ case and does not form a combination in order to cancel the longitudinal degree of freedom for the graviton.

The conclusion is that a generically modified 4D gravity $f(R,P,Q)$ has a ghost. One could argue that it is also possible to add a number of terms with
derivatives of curvature invariants, but that would just add extra ghost poles.  Correspondingly such modifications of gravity do not fit in the frame of
our current understanding of weakly coupled field theory. This ghost is a real problem and even in the particular case
$\left(f_P/2+2f_Q\right)|_{R=R_0}=0$, although there is no graviton ghost, there can be one in the scalar field. The best we can do is to ensure that
there is no ghost in the scalar field either. That would reduce this theory to scalar-tensor gravity which cannot accommodate both the acceleration of
the universe and solar system observations. On the other hand higher dimensional theories can lead to a consistent infrared modification of gravity, a
good example being the DGP model \cite{Deffayet:2001pu}, \cite{Dvali:2000hr}.

The authors would like to thank Gia Dvali for useful discussions.

\section*{Appendix A}
The inverse propagator is
\bb
&\;&O_{\alpha\beta\mu\nu}=
\left(Ag_{\alpha\mu}g_{\beta\nu}+
\left(B+CR \right)g_{\alpha\beta}g_{\mu\nu} \right)\nabla^4 -\nonumber\\
&-&2 \left( 
B+CR \right)g_{\mu\nu}\nabla^2\nabla_{\alpha}\nabla_{\beta} -
2A g_{\beta\nu}\nabla^2 \nabla_{\alpha}\nabla_{\mu}+\nonumber\\
&+&\left(A+B+CR\right)\nabla_{\alpha}\nabla_{\beta}\nabla_{\mu}\nabla_{\nu}+\nonumber\\
&+& \left(-\frac{3f_R}{8}+ \frac{R}{3}\left[-\frac{9f_P}{16}+f_{RR}\right]+
C\frac{R^2}{3}\right)g_{\alpha\beta}g_{\mu\nu}\nabla^2+\nonumber\\
&+&\frac{1}{2}\left(\frac{3f_R}{2}+R\left[\frac{5f_P}{2} +f_Q +\frac{f_{RR}}{3}\right]+
C\frac{R^2}{3}\right)g_{\mu\nu}\nabla_{\alpha}\nabla_{\beta}-\nonumber\\
&+& \frac{1}{8} \left(f - R f_R -\frac{R^2}{2}\left[\frac{f_P}{2}+\frac{f_Q}{3}-f_{RR}\right]+
C\frac{R^3}{2}\right)g_{\alpha\beta}g_{\mu\nu},
\ee
where
\bb
A&\equiv& \left[\frac{f_{P}}{2}+2 f_{Q}\right],\;\;\;\;
B\equiv \left[\frac{f_{P}}{2}+ f_{RR}\right],\nonumber\\
C&\equiv& \left[\frac{f_{PP}}{4}+\frac{f_{PQ}}{3}+\frac{f_{QQ}}{9}\right]R+\left[f_{RP}+\frac{2f_{RQ}}{3} \right],\nonumber\\
\nabla^2&\equiv& \nabla_{\alpha}\nabla^{\alpha},\;\;\;\; 
\nabla^4\equiv \nabla_{\alpha}\nabla^{\alpha}\nabla_{\beta}\nabla^{\beta}
\ee
Also, this expression is understood to be symmetrized  under the exchange of index pairs
$(\alpha\beta)\leftrightarrow(\mu\nu)$ and under the exchange of index inside the pairs,
$\alpha\leftrightarrow\beta$ and $\mu\leftrightarrow\nu$.

The covariant derivatives in this expression act on the second rank tensor $h_{\mu\nu}$.

\section*{Appendix B}
Below we give the expressions for the ten operators which span the space of solutions to the linearized 
field equations \cite{VanNieuwenhuizen:fi}: 
\bb
P^2&=&\frac12(\theta_{\mu\rho}\theta_{\nu\sigma}+\theta_{\mu\sigma}\theta_{\nu\rho})-\frac13\theta_{\mu\nu}\theta_{\rho\sigma},
\nonumber\\
P^1_m&=&\frac12(\theta_{\mu\rho}\omega_{\nu\sigma}+\theta_{\mu\sigma}\omega_{\nu\rho}+\theta_{\nu\rho}\omega_{\mu\sigma}+\theta_{\nu\sigma}\omega_{\mu\rho}),
\nonumber\\
P^1_e&=&\frac12(\theta_{\mu\rho}\omega_{\nu\sigma}-\theta_{\mu\sigma}\omega_{\nu\rho}-\theta_{\nu\rho}\omega_{\mu\sigma}+\theta_{\nu\sigma}\omega_{\mu\rho}),
\nonumber\\
P^1_b&=&\frac12(\theta_{\mu\rho}\theta_{\nu\sigma}-\theta_{\mu\sigma}\theta_{\nu\rho}),
\nonumber\\
P^1_{me}&=&\frac12(\theta_{\mu\rho}\omega_{\nu\sigma}-\theta_{\mu\sigma}\omega_{\nu\rho}+\theta_{\nu\rho}\omega_{\mu\sigma}-\theta_{\nu\sigma}\omega_{\mu\rho}),
\nonumber\\
P^1_{em}&=&\frac12(\theta_{\mu\rho}\omega_{\nu\sigma}+\theta_{\mu\sigma}\omega_{\nu\rho}-\theta_{\nu\rho}\omega_{\mu\sigma}-\theta_{\nu\sigma}\omega_{\mu\rho}),
\nonumber\\
P^0_s&=&\frac13\theta_{\mu\nu}\theta_{\rho\sigma},
\nonumber\\
P^0_w&=&\omega_{\mu\nu}\omega_{\rho\sigma},
\nonumber\\
P^0_{sw}&=&\frac1{\sqrt3}\theta_{\mu\nu}\omega_{\rho\sigma},
\nonumber\\
P^0_{ws}&=&\frac1{\sqrt3}\omega_{\mu\nu}\theta_{\rho\sigma},
\nonumber
\ee
where the transversal and longitudinal projectors in the momentum space are respectively
\bb
\theta_{\mu\nu}=\delta_{\mu\nu}-\frac{\partial_\mu \partial_\nu}{\partial^2},\qquad \omega_{\mu\nu}=\frac{\partial_\mu \partial_\nu}{\partial^2}.
\nonumber
\ee

%%%%%%%%%%%%%%%%%%%%%%

\end{document}